# Bunch Length Diagnostics: Current Status and Future Directions

*Allan Gillespie*
University of Dundee, Scotland

**Abstract**
Charged particle bunch length detection is one of the most challenging measurements in particle and accelerator physics, especially when the bunch length reduces below tens of femtoseconds. Since these measurements are most critical in high energy electron accelerators and devices with ultrashort electron bunches—such as the plasma wakefield and related accelerators—this discussion is limited to electron bunch detection, although the results clearly have significance for any ultra-relativistic particle bunches.

## 1 Introduction

Perhaps the fundamental question is: why do we need femtosecond electron bunch profile measurements? The need arises because a new generation of electron accelerators is pushing the limits of ultrashort bunches for a variety of practical reasons. The latest generation of free electron lasers (FELs) require kA peak currents for high collective gain, corresponding to bunch charges of 10–100 pC within 10 fs FWHM. In the future, even ultra-low-emittance storage rings will require very short electron bunches, especially when they incorporate FEL sections. Projected particle physics colliders ($e^+$-$e^-$ and others) such as CLIC and the ILC will require short bunches with high charge, high quality beams to maximize luminosity. For example, the CLIC concept calls for 150 fs electron and positron colliding beams with precise demands on monitoring. Finally, laser plasma wakefield accelerators and their many variants can only operate via ultrashort electron bunches of 1–5 fs FWHM, and these numbers will continue to decrease.

It is therefore clear that the measurement of such ultrashort bunches poses a fundamental challenge, and that novel methods will be required, especially for the relatively unstable beams currently generated by the latest ultrashort-bunch devices. Note that in this discussion we interpret 'sub-picosecond bunch length monitor' to mean an accelerator diagnostic that attempts to provide a 'true' temporal profile of an ultrafast electron bunch, rather than one which only aims to inform on the first moments of the bunch profile, or on the presence of structure on a particular time scale of interest.

It is also worth noting that many effects—such as wake fields, space charge, coherent synchrotron radiation (CSR), and collective instabilities, along with accelerator stability, jitter, and drift—can have a significant influence on the accelerator bunch profile. It follows that we must therefore develop **single-shot diagnostics** for such ultrashort bunches.

## 2      Phase space transformations

Most charged-particle beam diagnostics are carried out by performing phase space transformations of the charged particles themselves or their products (usually photons).

We are working within a 6-dimensional particle phase space (Fig. 1), where the usual horizontal and vertical dimensions ( x, x′ (or θ), y and y′ (or ϕ) ) are augmented by the momentum deviation δ and time t (or equivalently the bunch length l ). The *longitudinal phase space* is the projection on these last two dimensions, and in particular the time, which is very difficult to measure with femtosecond accuracy. No diagnostics exist for the entire distribution.

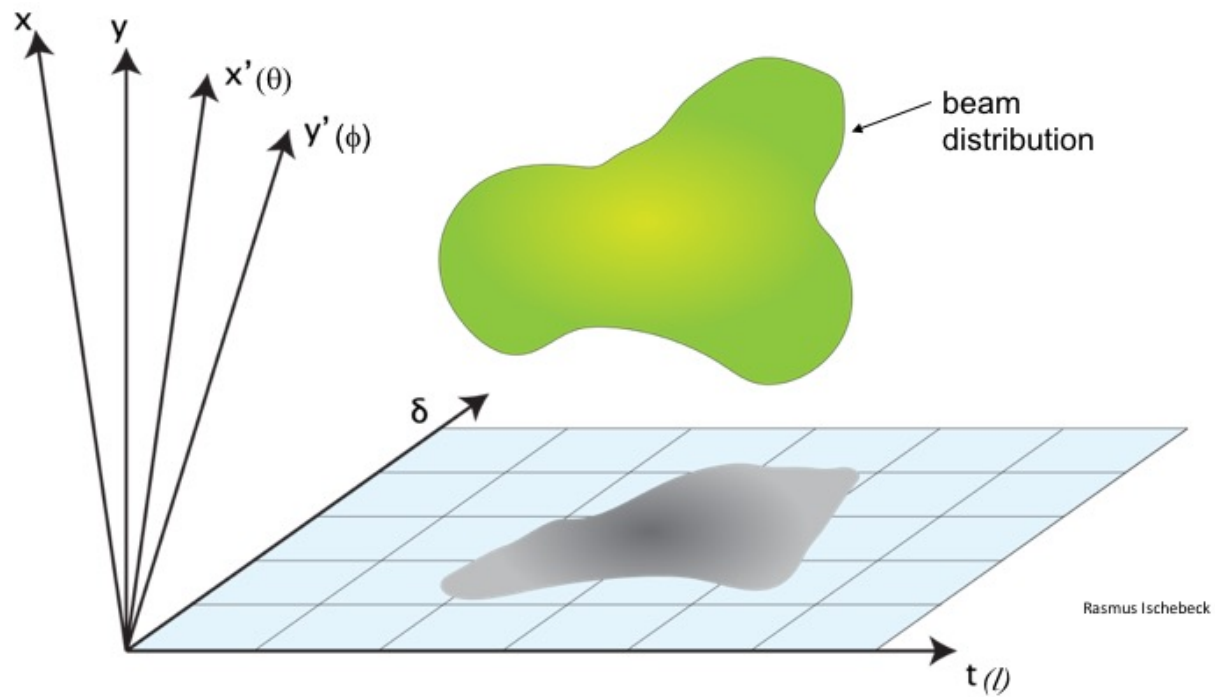

**Fig. 1:** Six-dimensional phase space

In general, charged-particle phase space transformations can be performed by:

i)   quadrupole magnets       (x-x′, y-y′);
ii)  dipole magnets            (δ-x′);
iii) RF deflectors             (t-x′).

Specialist transformations can be carried out by, for example:

i)   off-crest RF acceleration;
ii)  dispersive systems;
iii) magnetic bunch compressors;
iv)  energy compressor systems.

## 2.1 Classes of diagnostics

There exist two distinct classes of longitudinal diagnostics, which can be grouped by their similar physics and capabilities (or limitations).

1. Direct particle techniques
$$\rho(t) \rightarrow \rho(x).$$

   Here, the longitudinal bunch extent is converted to transverse imaging.

   Examples include RF zero-phasing and transverse deflecting cavities.

2. 'Radiative' techniques
$$\rho(t) \rightarrow E(t),$$

   where the bunch charge is induced to radiate—involving both propagating and non-propagating fields.

   This category includes both spectral domain and time domain techniques.

### *2.1.1 Class 1: direct particle techniques*

Here we act *directly* on the electron beam to deflect or disperse the beam particles. We shall see that in all cases we have to be very sure about the electron beam optics throughout the diagnostic.

An example of this technique applied to *photons* is the RF streak camera, a traditional method used to determine photon pulse length. The time variation of the incoming photon pulse intensity—often a replica of the longitudinal bunch profile—is converted into a spatial intensity variation using a photocathode followed by an RF sweep circuit that displays the transverse sweep on a phosphor screen. Such cameras (and related synchroscan streak cameras) can achieve a sub-picosecond time resolution [1].

In the RF zero-phasing technique, the bunch distribution is passed through an RF cavity centred on zero phase (no acceleration), introducing an energy chirp into the bunch by virtue of the nearly-linear RF field. A downstream magnetic spectrometer is then used to convert this chirp into a transverse bunch profile. The resulting time resolution is dependent on the gradient of energy gain, the spectrometer dispersion (which needs to be high), and the initial energy spread (assumed to be low), and it is assumed that there is no initial γ–z correlation present in the incoming beam.

The fundamental disadvantage of this technique is that it is destructive to the electron beam, and hence one cannot utilize the detected electron bunch downstream.

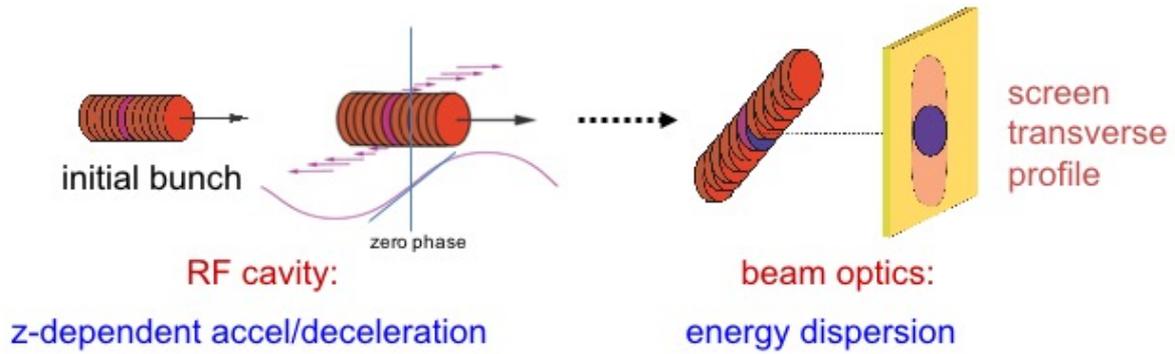

**Fig. 2:** RF zero phasing

A good example of this method is the SLAC LCLS system operated at 4.7 GeV [2] which delivered an approximate 1 fs rms bunch length by making use of 550 m of linac at RF zero crossing and the 6 m dispersion of the SLAC A-line spectrometer.

The classic technique used to measure electron bunch length involves the use of a **transverse deflecting structure.** Here, a specialized RF cavity designed to provide hybrid deflecting modes $HEM_{1,1}$ (often called a LOLA cavity after its originators at SLAC) produces a large time-varying transverse electric field to 'streak' the beam across a downstream monitor screen (Fig. 3). This can therefore be looked at as a sort of intra-beam streak camera. Such 3 m-long S-band 2856 MHz structures have been installed on the SLAC LCLS linac, but they are invasive to operation for photon users of LCLS since they are naturally destructive to the electron beam.

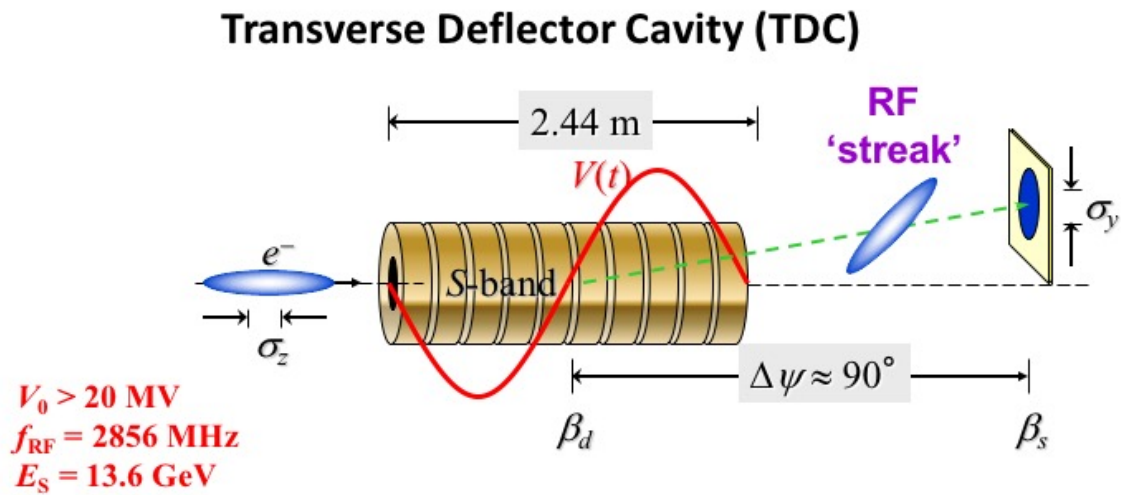

$$\sigma_y^2 = \sigma_{y0}^2 + \beta_d \beta_s \sigma_z^2 \left( \frac{k_{RF} e V_0}{E_s} \sin \Delta\psi \cos\phi \right)^2$$

- Map time axis on to transverse coordinate
- Simple calibration by scan of cavity phase

**Fig. 3:** TDC principle

However, for applications involving the free-electron laser (FEL) where the electrons are dumped and the undulator photons proceed to the experimental areas, this technique can provide unmatched time resolution exceeding 1 femtosecond, so far unreachable by other techniques. This is achievable at SLAC by using an X-band transverse deflecting cavity (XTCAV) operating at 11.424 GHz (Figs. 4 and 5). The horizonal undulator is followed by the X-band RF deflector, which generates a horizontal streak. This is then passed to a high-resolution vertical dipole producing a vertical energy dispersion on the monitor screen. The overall effect is to display an energy–time plot which gives critical information on the FEL operation and allows calculation of the resulting spectrum of hard X-rays [3,4,5].

X-band structures of this type have also been installed at other electron accelerators, including FLASH at Hamburg.

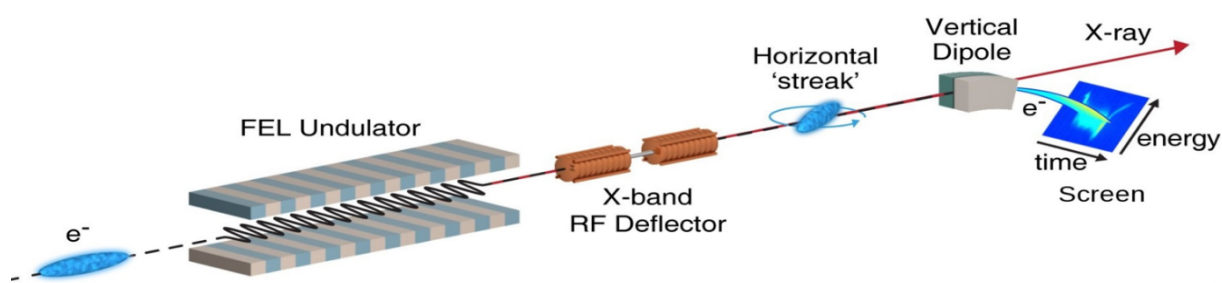

**Fig. 4:** Principle of XTCAV transverse deflecting cavity at SLAC

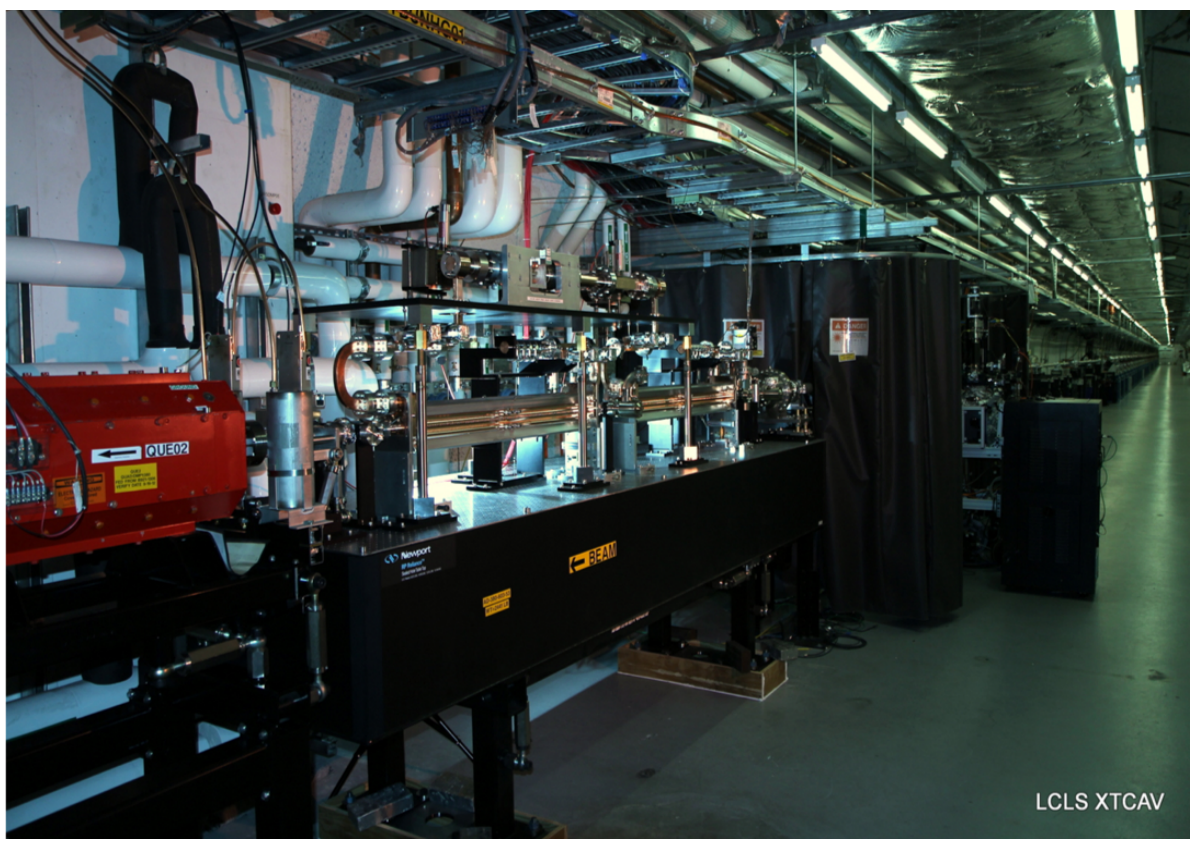

**Fig. 5:** Photo of the LCLS XTCAV at SLAC

Operating at a frequency of 11.424 GHz gives 8 times better temporal resolution over the original structures and allows continuous non-invasive single-shot photon operation at 120 Hz. Figure 6 shows a typical measured bunch profile. The demonstrated time resolution of 0.8 ± 0.2 fs is currently unmatched by any other technique.

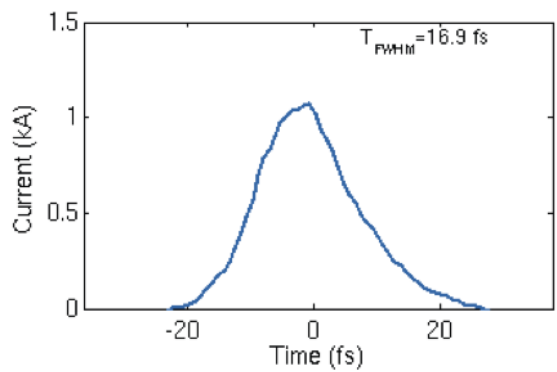

**Fig. 6:** Measured bunch profile from XTCAV

The usefulness of the XTCAV in measuring the FEL spectrum—with and without the LCLS FEL operation—is shown in Fig. 7. The energy loss and spread due to transfer of energy from electrons to photons is clearly visible.

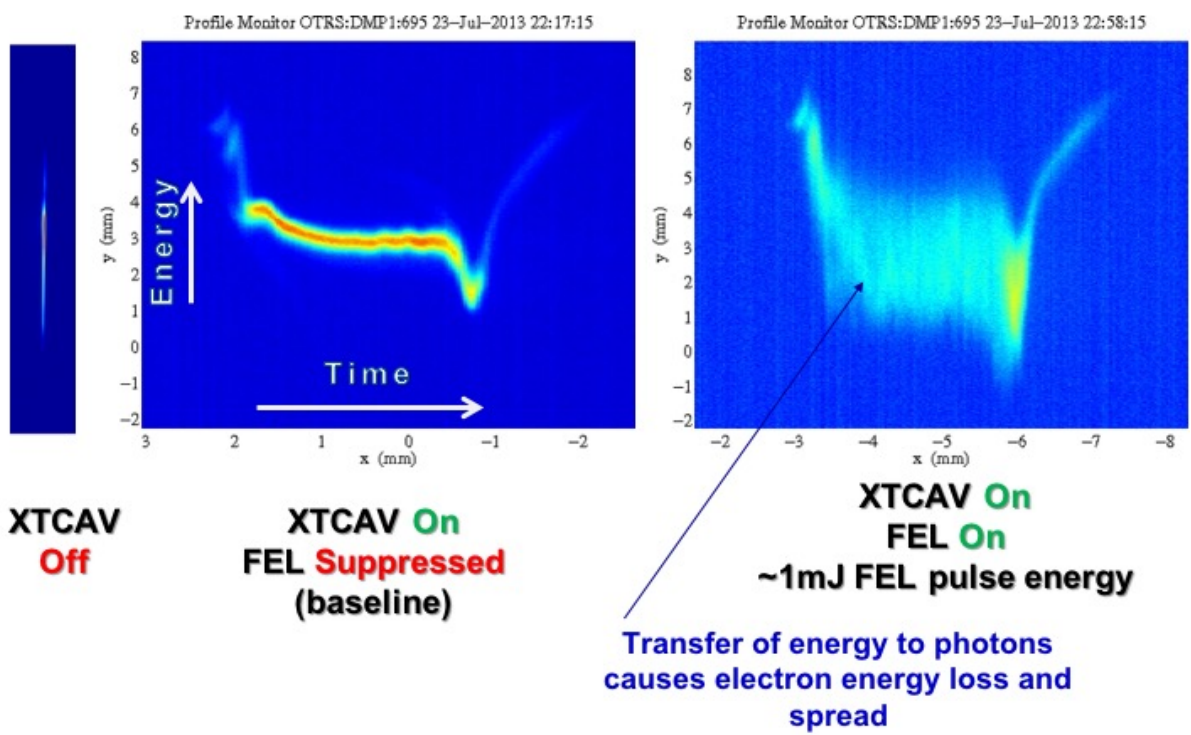

**Fig. 7:** FEL electron spectrum produced by XTCAV

It is worth noting that proper behaviour depends on optimized beam transport optics between TDC and view screen, and that the set-up that gives optimum performance and resolution may not be the same as the one required for normal accelerator/FEL operation. The TDC can also give access to slice parameters of electron beam, important for optimizing the performance of an FEL. It does, however, sacrifice 1 bunch and therefore if there exists a requirement for *absolutely* non-intercepting operation within an accelerator, this may make a TDC impossible.

### 2.1.2    Class 2: 'radiative' techniques

Here we cause the Coulomb field of the electron bunch to radiate in a controlled manner, and subsequently infer the bunch profile from the emitted radiation spectrum. The general methodology is to cause the bunch to radiate coherently, propagate the radiation to the observation point, then measure the spectrum i.e., the intensity–time profile.

The problem here is that we cannot normally record the phase information, and this makes inferring the charge density from the spectral profile a more difficult task. This is not made easier by attenuation, dispersion, and diffraction—with the longer wavelengths significantly affected by diffraction, and therefore at least partially missing in the reconstruction process.

It is important to realize that these limitations affect *all* versions of the radiative technique, including coherent diffraction radiation (CDR), coherent transition radiation (CTR), Smith-Purcell radiation, coherent synchrotron radiation (CSR), etc.

Remember also that the field radiated or probed is related to the Coulomb field *near* the electron bunch. Therefore, the time response δt and spectrum of the detected field is dependent on the spatial deviation r from the bunch axis.

$$\delta t \sim 2r/c\gamma \,.$$

This demonstrates that ultrafast time resolution requires close proximity to the bunch. It is also worth noting that this is one of the few detection techniques which improves as the beam energy γ increases. One can easily show that a 20 fs time resolution is only obtainable for beam energies above 1 GeV.

## 3    Coherent radiation

For wavelengths *shorter than* the bunch length, the particles within the bunch radiate incoherently, with the power emitted proportional to the number $N$ of particles. However, for wavelengths *equal to or longer than* the bunch length, particles emit radiation *coherently* with the emitted power dependent on the bunch length and scaling as the *square* of the number of particles.

$$S(\omega) = S_p(\omega)\left[N + N(N-1)F(\omega)\right],$$

where $S(\omega)$ represents the detected radiation spectrum, $S_p(\omega)$ the single particle spectrum, $N$ the number of particles in the bunch, and $F(\omega)$ the longitudinal bunch form factor—which is the modulus squared of the Fourier transform of the longitudinal particle distribution $\rho(s)$.

Measuring the power spectrum therefore allows the form factor to be calculated, from which an indirect measurement of the bunch length is possible.

What is obtained, however, is the form factor rather than the longitudinal distribution. In order to reconstruct the longitudinal bunch profile it is necessary to perform an inverse-Fourier transformation using phase recovery algorithms making use of the Kramers–Kronig relation.

Figure 8 illustrates the most common spectral domain radiative techniques [6].

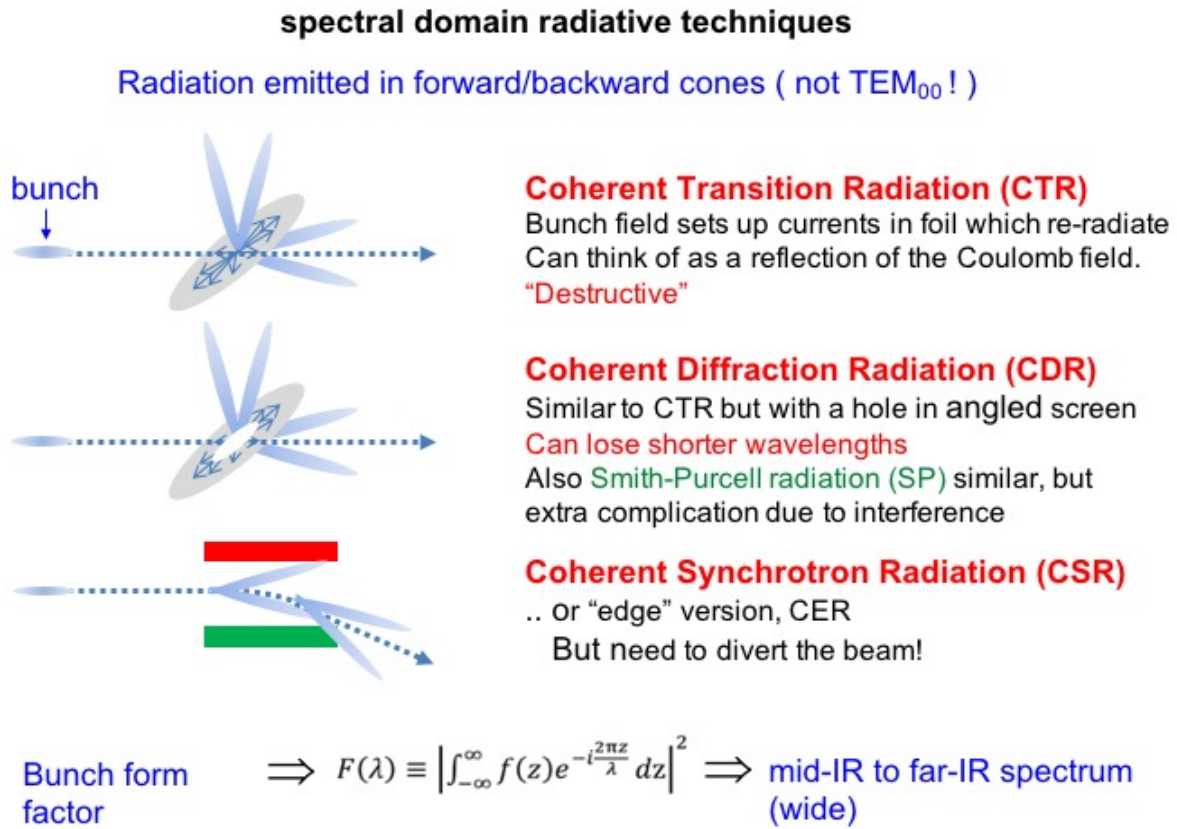

**Fig. 8:** Spectral domain radiative techniques

Coherent transition radiation (CTR) detection typically involves placing a thin (metal) foil at 45° to the beam axis (Fig. 7), causing the twin-lobed radiation to be emitted in both the forward and 'reflection' directions [7]. Detection at 90° is convenient, capturing as much of the double lobe as feasible. The radiation arises when a charge traverses a boundary between media with differing dielectric constants, and can be understood in terms of image charges. The sudden disappearing and appearing surface charge distribution as the charge crosses the boundary leads to radiation on an emtosecond time scale. Provide that the electron beam energy is not too low (say above 10 MeV) it is possible to capture the full backward TR angular distribution, generating a very useful beam screen or more detailed diagnostic. One possible drawback is that this is a partially intercepting device.

Coherent diffraction radiation (CDR) generates a similar double-lobed distribution by passing the beam through a hole in a metal foil or past the edge of a similar foil [8]. Again, both backward and forward CDR is generated.

Smith-Purcell radiation [9,10] is produced when a charged particle beam interacts close to a periodic metallic structure parallel to the beam axis, generating a wide angular dispersion of wavelengths that must be captured by a series of angular detectors. Coherent enhancement occurs for bunch lengths shorter than, or equal to, the emitted wavelengths. The bunch profile is not explicitly determined from the data; instead, experimental spectra are compared with calculated spectra based on trial bunch profiles. Typically, sub-picosecond bunch lengths are measurable.

Finally, coherent synchrotron radiation (CSR) is generated when the beam is bent in a dipole magnet and is widely used as a diagnostic in storage rings, especially to measure very small beam *sizes*. The edge of such dipole fields can also give rise to coherent edge radiation (CER) with similar properties and limitations.

All of these radiative techniques are subject to the power spectrum limitation mentioned above.

# 4 Electro-optic methods of bunch distribution measurement

For suitably relativistic beam energies the Coulomb field of each electron flattens out transverse to its direction of motion, becoming more disc-like, resulting in the overall Coulomb field of the bunch becoming an accurate representation of the charge distribution. This makes it possible to devise techniques that accurately sample this transverse Coulomb field very close to the bunch axis.

For example, one can use ultrafast laser sampling and electro-optic crystals to 'copy' the Coulomb field distribution of the electron bunch on to a synchronized pulsed laser beam. A range of standard techniques in ultrafast optics are then used to 'decode' the encoded laser pulse, extracting a 'true' time-resolved longitudinal beam bunch distribution [11]. In principle, this can yield a useful non-intercepting bunch profile monitor.

The basic electro-optic (EO) diagnostic principle is shown schematically in Fig. 9:

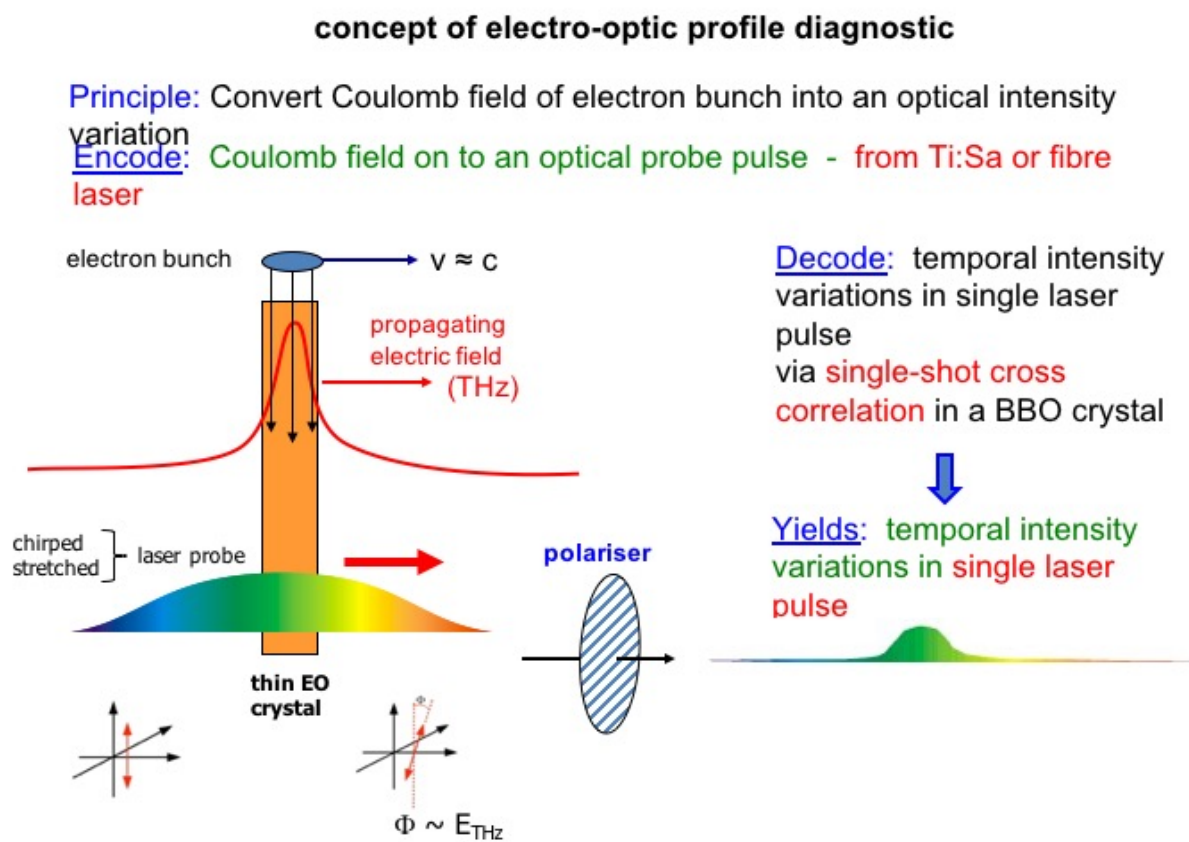

**Fig. 9:** Concept of electro-optic (EO) diagnostic

The bunch Coulomb field is allowed to sweep through a thin EO crystal placed adjacent to the bunch axis. In the simplest interpretation this field renders the crystal birefringent, which induces a temporal modulation in the intensity of a chirped probe laser passed through the same crystal. A shift in the polarization of the chirped beam allows this modulation to be decoded.

The simplest decoding technique, spectral decoding (SD), utilizes the time–wavelength correlation within the input probe laser pulse, and a subsequent spectral measurement of the probe, to unfold the temporal profile of the Coulomb field [12]. This use of a spectral measurement, and the inherent inseparability of temporal and spectral variations for ultrashort pulses, ultimately prevents the application of this technique to very short bunch characterization. The SD technique is therefore limited to electron bunch lengths greater than one picosecond.

There exist a wide range of electro-optic techniques, summarized in Fig. 10 [13,14,15]. For brevity, only a typical example will be considered here in detail—electro-optic temporal decoding (EOTD), which currently exhibits the highest time resolution.

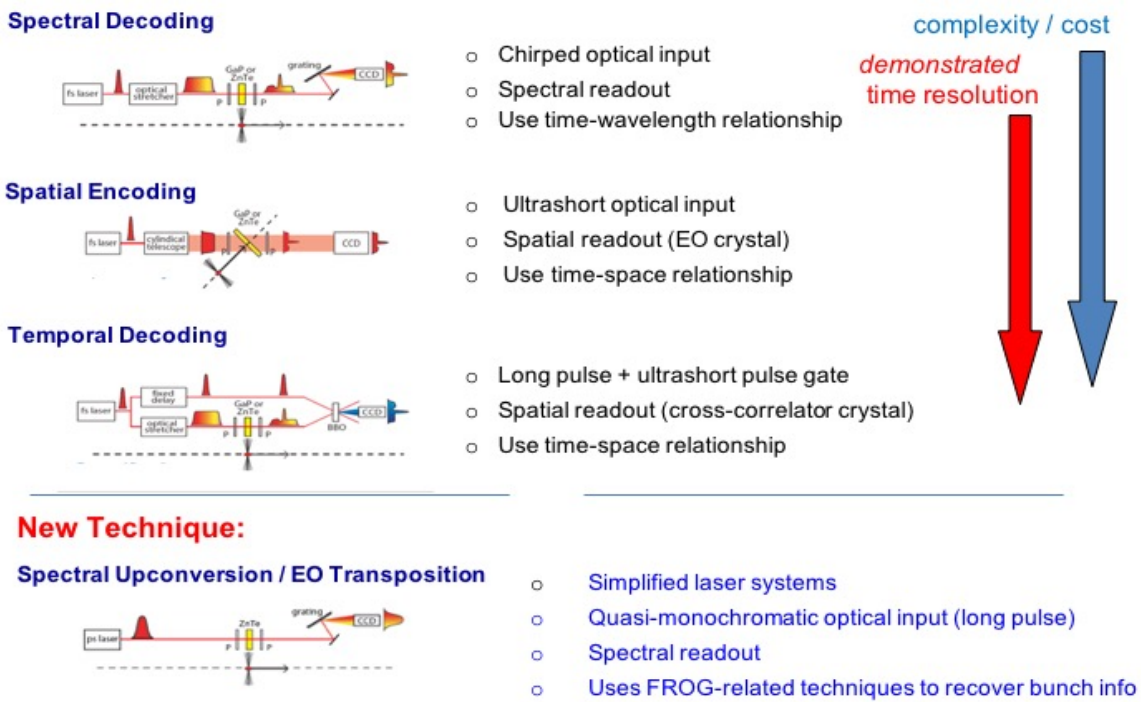

**Fig. 10:** Range of EO techniques

The temporal decoding technique measures the induced intensity modulation directly in the time domain through a time-to-space mapping. This is done external to the accelerator beamline in an optical cross-correlator using second harmonic generation (SHG) in a BBO crystal (Fig. 11) which converts the temporal profile of the probe pulse into a spatial image of the SHG pulse.

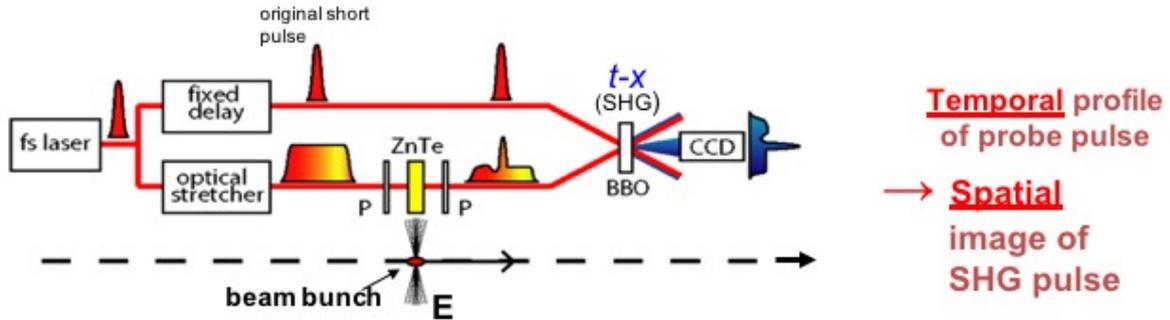

**Fig. 11:** Single-shot EOTD

The principal shortcoming of the EOTD technique lies in the frequency response of the electro-optic crystals employed. Due to phonon resonances within these inorganic crystals (typically GaP or ZnTe) the high frequency response rolls off at around 10 THz for thin samples (Fig. 12) whereas one would like a response flat to beyond 30–40 THz. This may become available via organic crystals such as DAST but there are severe materials problems with such samples. Some work has been done to manufacture 'metamaterials' with the desired frequency response [16] but so far no ideal material has been produced.

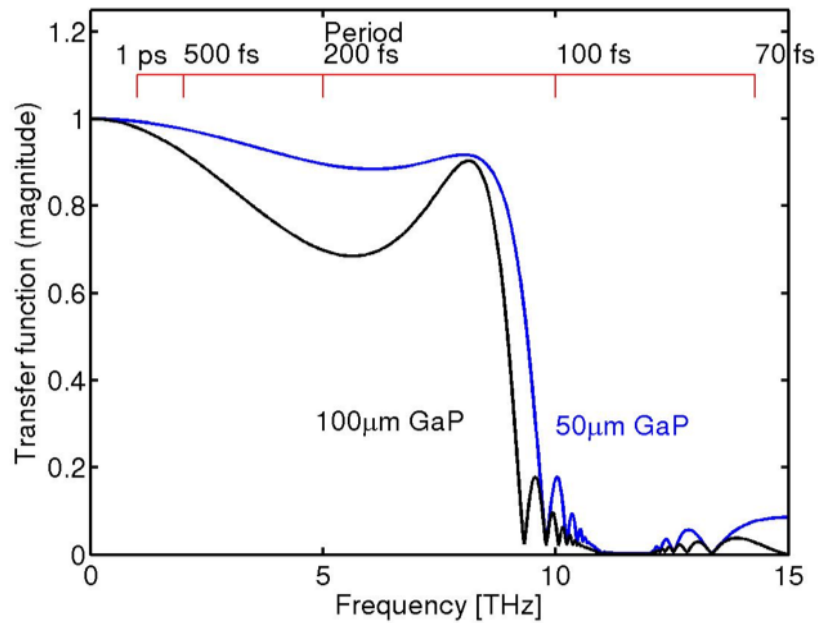

**Fig. 12:** GaP frequency response

The EO detection concept is traditionally described via an effective refractive index change induced by the Coulomb field. An alternative description [17] based on the non-linear frequency mixing of Coulomb and optical fields allows for the analysis to be extended to chirped optical probe pulses and ultrashort Coulomb fields in a rigorous and straightforward manner. Based on this analysis, electro-optic diagnostics with a target time resolution of 20 fs rms, and with intrinsically improved stability and reliability, are under development for the CLIC linear collider [18].

The new system is based on 'spectral upconversion', where a quasi-CW laser probe is mixed with the sub-picosecond duration Coulomb field pulse of the relativistic electron bunch. The electro-optic effect carves an 'optical-replica' of the longitudinal charge distribution from the narrow-bandwidth probe, simultaneously up-converting the bunch spectrum to optical frequencies. By using frequency resolved optical gating (FROG), an extension of autocorrelation, the optical replica can then be characterized on a femtosecond time scale. This scheme, termed electro-optic transposition (EOT), therefore bypasses the requirement for unreliable femtosecond laser systems. The high pulse energy required for single-shot pulse measurement via FROG is produced through optical parametric amplification of the optical-replica pulses. The complete system is based on a single nanosecond-pulse laser—resulting in a reliable system with greatly relaxed timing requirements.

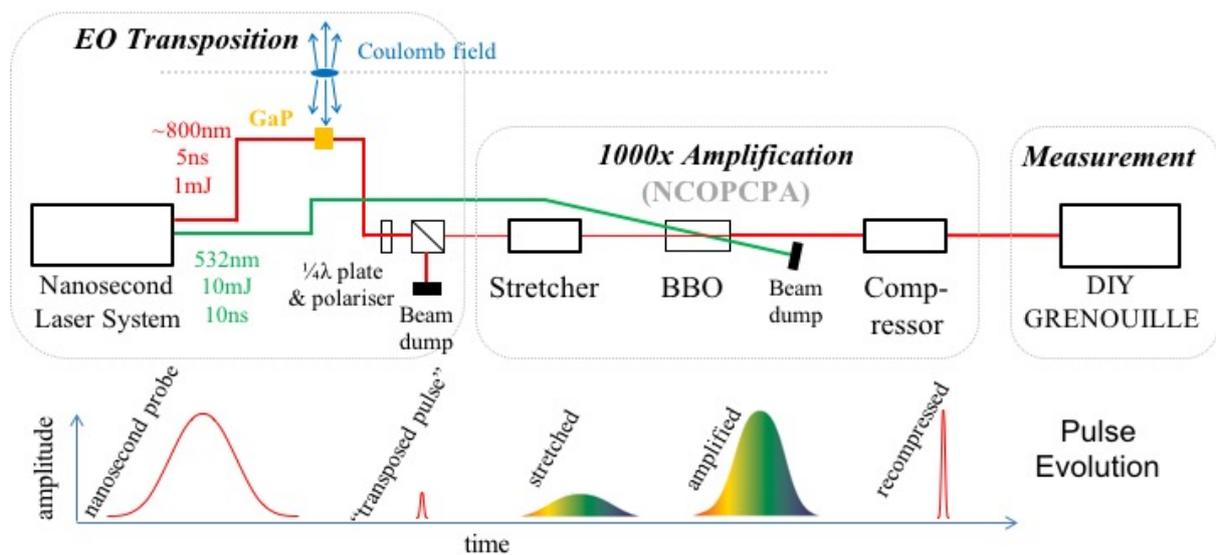

**Fig. 13:** Principle of electro-optic transposition (EOT)

The use of a nanosecond laser brings reliability, and the full spectral amplitude and phase are measured via the FROG technique. The Coulomb field (bunch profile) is then calculated via time-reversed propagation of the detected pulse. When fully developed, such techniques offer a novel solution to ultrashort bunch measurements.

There exist a range of other diagnostic methods [19–24] including radiative and exotic techniques which we do not have space to mention here. This is a very active field, and one which will continue to play a key role in future accelerator operations.

**Acknowledgements**

We would like to acknowledge contributions of graphics, photos and plots provided by:

    Rasmus Ischebeck, PSI Villigen
    David Walsh, STFC Daresbury Laboratory
    Patrick Krejcik, SLAC National Accelerator Laboratory
    Bernd Steffen and Christopher Gerth, E-XFEL, Hamburg

# References

[1] Y. Otake *et al.*, *Phys. Rev. Spec. Top. Accel. Beams* **16**, 042802 (2013), https://doi.org/10.1103/PhysRevSTAB.16.042802.

[2] Z. Huang *et al.*, *Phys. Rev. Spec. Top. Accel. Beams* 13 (2010), 092801, https://doi.org/10.1103/PhysRevSTAB.13.092801.

[3] P. Krejcik *et al.*, Proceedings of IBIC2013, Oxford UK, 308-311 (2013), http://accelconf.web.cern.ch/AccelConf/IBIC2013/papers/tual2.pdf.

[4] Y. Ding *et al.*, SLAC-PUB-16105, http://www.slac.stanford.edu/pubs/slacpubs/16000/slac-pub-16105.pdf.

[5] Y. Ding *et al.*, *Phys. Rev. Spec. Top. Accel. Beams* **14** (2011) 120701, https://doi.org/10.1103/PhysRevSTAB.14.120701.

[6] W.A. Gillespie *et al.*, Proceedings of IPAC 2015, Richmond, VA, USA (2015), transparencies, http://accelconf.web.cern.ch/AccelConf/IPAC2015/html/auth1204.htm.

[7] I. Nozawa *et al.*, *Phys. Rev. Spec. Top. Accel. Beams* **17** (2014) 7, 072804, https://doi.org/10.1103/PhysRevSTAB.17.072803.

[8] M. Castellano *et al.*, *Phys. Rev. E* **63** (2001) 056501, https://doi.org/10.1103/PhysRevE.63.056501.

[9] G. Doucas *et al.*, *Phys. Rev. Lett.* **69** (1992) 176, https://doi.org/10.1103/PhysRevLett.69.1761.

[10] K. Ishi *et al.*, *Phys. Rev. E* **51** (1995) R521, https://doi.org/10.1103/PhysRevE.51.R5212.

[11] I. Wilke *et al.*, *NIMA* **483** (2002) 282–285, https://doi.org/10.1016/S0168-9002(02)00328-5.

[12] S.P. Jamison *et al.*, *Optics Letters* **31** (2006) 1753, https://doi.org/10.1364/OL.31.001753.

[13] A.L. Cavalieri *et al.*, *Phys. Rev. Lett.* **94** (2005) 11480, https://doi.org/10.1103/PhysRevLett.94.114801.

[14] I. Wilke *et al.*, *Phys. Re. Lett.* **88** (2002) 124801, https://doi.org/10.1103/PhysRevLett.88.124801.

[15] G. Berden *et al.*, *Phys. Rev. Lett*. **99** (2007) 164801, https://doi.org/10.1103/PhysRevLett.99.164801.

[16] S.A. Zolotovskaya *et al.*, *Nanotechnology* **27** (2016) 435703, https://doi.org/10.1088/0957-4484/27/43/435703.

[17] S.P. Jamison *et al.*, *Applied Phys. Lett.* **96** (2010) 231114-231114-3, https://doi.org/10.1063/1.3449132.

[18] D.A. Walsh *et al.*, Proceedings of IBIC2016, Barcelona, Spain, 752-755, http://dx.doi.org/10.18429/JACoW-IBIC2016-WEPG49.

[19] M. Castellano *et al.*, *Phys. Rev. E.* **63** (056501) (2001), https://doi.org/10.1103/PhysRevB.63.060102.

[20] B. Feng *et al.*, *Nucl. Instr. Meth. Phys. Res. A* **475** (2001) 492–497, https://doi.org/10.1016/S0168-9002(01)01605-9.

[21] T.J. Maxwell *et al.*, *Phys. Rev. Lett.* **111** (2013) 18, 184801, https://doi.org/10.1103/PhysRevLett.111.184801.

[22] A. Aryshev *et al.*, *J. Phys. Conference Series* **236** (2010) 012008, https://doi.org/10.1088/1742-6596/236/1/012008.

[23] M.H. Helle *et al.*, *Phys. Rev. ST Accel. Beams* **15** (2012) 052801, https://doi.org/10.1103/PhysRevSTAB.15.052801.

[24] I. Dornmair *et al.*, *Phys. Rev. Accel. Beams* **19** (2016) 062801, https://doi.org/10.1103/PhysRevAccelBeams.19.062801.